\documentclass[a4paper,12pt]{article}

\usepackage[utf8x]{inputenc}
\usepackage[T1]{fontenc}
\usepackage{times}
\usepackage{geometry}
\usepackage{titlesec}
\titleformat{\section}
{\bfseries\uppercase}{\thesection.}{1em}{}
\titleformat{\subsection}
{\bfseries}{\thesection.\thesubsection.}{1em}{}

\usepackage{graphicx} 
\usepackage{multirow} 
\usepackage{cite}
\usepackage{breakurl}
\usepackage{indentfirst}
\usepackage{amsmath, amssymb, amsfonts, bm}
\usepackage{txfonts}
\usepackage{enumitem}
\usepackage{xcolor}
\usepackage{enumitem}
\usepackage{subcaption}

\titlespacing*{\subsection}{0pt}{1.5em}{0.2em}
\titlespacing*{\section}{0pt}{1.5em}{0.2em}

\renewcommand\eqref[1]{Equation~\ref{#1}}

\renewcommand{\thesection}{\arabic{section}}
\renewcommand{\thesubsection}{\arabic{subsection}}

\makeatletter
\renewcommand\@biblabel[1]{#1.}
\makeatother

\newlength{\bibitemsep}
\newlength{\bibparskip}
\let\oldthebibliography\thebibliography
\renewcommand\thebibliography[1]{%
  \oldthebibliography{#1}%
  \setlength{\parskip}{\bibitemsep}%
  \setlength{\itemsep}{\bibparskip}%
}
 
\begin{document}

\begin{center}
	\includegraphics[width=3.50in]{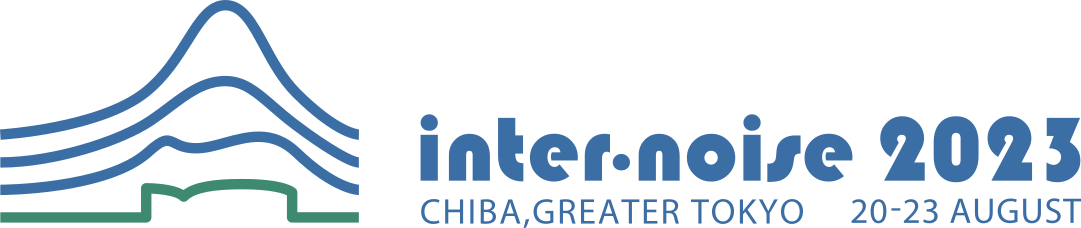}
\end{center}
\vskip.5cm

\begin{flushleft}
\fontsize{16}{20}\selectfont\bfseries
A Computation-efficient Online Secondary Path Modeling Technique for Modified FXLMS Algorithm
\end{flushleft}
\vskip1cm

\renewcommand\baselinestretch{1}
\begin{flushleft}

Junwei Ji\footnote{JUNWEI002@e.ntu.edu.sg}, Dongyuan Shi\footnote{dongyuan.shi@ntu.edu.sg}, Woon-Seng Gan\footnote{ewsgan@ntu.edu.sg}, Xiaoyi Shen\footnote{xiaoyi.shen@ntu.edu.sg}, Zhengding Luo\footnote{LUOZ0021@e.ntu.edu.sg}\\
School of Electrical and Electronic Engineering, Nanyang Technological University\\
50 Nanyang Avenue, Singapore, 639798\\


\end{flushleft}
\textbf{\centerline{ABSTRACT}}\\
\textit{This paper proposes an online secondary path modelling (SPM) technique to improve the performance of the modified filtered reference Least Mean Square (FXLMS) algorithm. It can effectively respond to a time-varying secondary path, which refers to the path from a secondary source to an error sensor. Unlike traditional methods, the proposed approach switches modes between adaptive ANC and online SPM, eliminating the use of destabilizing components such as auxiliary white noise or additional filters, which can negatively impact the complexity, stability, and noise reduction performance of the ANC system. The system operates in adaptive ANC mode until divergence is detected due to secondary path changes. At this moment, it switches to SPM mode until the path is remodeled and then returns to ANC mode. Furthermore, numerical simulations in the paper demonstrate that the proposed online technique effectively copes with the secondary path variations.
} 

\section{INTRODUCTION}
\label{sec:intro}
\noindent
The technique of active noise control (ANC) is intended to attenuate unwanted noise by generating anti-noise, which can be obtained through some electronic and electro-acoustic components combined with advanced algorithms \cite{KuoANC}. In these two decades, many strategies for ANC performance improvement are proposed to address some practical issues \cite{LamTenQues, lu2021survey, kajikawa2012recent}, such as output constraint to deal with signal distortion \cite{SHIconstrain, Shifrequency, xiao2020ultra, liu2019wireless}, wireless techniques to enhance the signal-to-noise ratio (SNR) of the reference signal \cite{Shenwireless2022, Shenwirelessjsv}, multichannel ANC to enlarge the quiet area \cite{shi2020active1, IWAI2019151, zhang2021spatial, sun2022secondary} and solutions for high computational cost in multichannel application \cite{Ferrerdistributed, Chendistributed, Chudistributed, Lidistributed}, etc. In the era of artificial intelligence (AI), some ANC research based on deep learning also springs up \cite{Shitranfercnn, Luohybriddeep, zhang2021deep,zhang2023deep, shi2023active}.

Among adaptive control algorithms~\cite{lu2021surveyII,shi2023active,lu2021survey,yang2019mean,ferrer2010transient}, the filtered reference least mean square (FXLMS) algorithm \cite{KuoANC,yang2020stochastic}, one of the most realizable methods, is still widely used nowadays. It is a derivative of the least mean square (LMS) algorithm proposed to compensate for the delay caused by the secondary path, which refers to the path from a secondary source to an error sensor, including necessary electronic components and acoustic path \cite{KuoANC}. However, in the algorithm, the reference signal is filtered by the secondary path estimate before being fed into the LMS algorithm, resulting in a slower convergence speed than the conventional LMS~\cite{ElliottANC}.
Therefore, the modified FXLMS algorithm removes the effect of the secondary path to increase the convergence speed at the cost of the computational burden \cite{BjarnasonMFXLMS}. Nevertheless, the performance of modified FXLMS also depends on the accuracy of secondary path modelling \cite{BjarAnalyfxlms, LopesBehMFXLMS}. Hence, it is necessary to estimate the time-varying secondary path during the noise cancellation. One of the most popular solutions is the online secondary path modelling technique.

In terms of online secondary path modeling, the approach of injecting random white noise into the ANC system to accomplish secondary path identification \cite{ErrikssonOSPM} is commonly used and proved to have good performance \cite{BaoComparOSPM}. In addition, Zhang proposed a cross-updated method \cite{ZhangXupdated} based on Bao's structure \cite{BaoJSV} to further improve the performance of Eriksson's idea on online SPM with additional random noise \cite{ErrikssonOSPM}. Furthermore, they introduce a different adaptive filter together with two other adaptive filters, one for ANC and another one for SPM, to reduce the perturbation caused by disturbance to modeling the secondary path and suppress the effect caused by the injected noise to the control filter. Aside from these modifications, they publish an "auxiliary noise power scheduling" strategy to relieve the effect on ANC performance owing to the additional noise introduced to the system \cite{HuiSPM, Zhangpowerscheduling}. However, this additional adaptive filter raises the complexity of the overall system. Hence, Akhtar \cite{AkhtarVSSSPM} suggested modeling the secondary path using a variable step size (VSS) LMS algorithm instead of introducing an additional adaptive filter. Its improved versions are also developed to enhance ANC performance \cite{Akhtarschdeduling, DavariSPM}. However, its computation complexity is still relatively high. In recent years, stage-based approaches to addressing this issue have been presented in \cite{ShenAltern, Kimtwostage}. In \cite{Pradhan5stage}, it employs a 5-stage method in which it first estimates the primary path, then initializes the controller with a single gain to generate the control signal for estimating the secondary path, and lastly initiates ANC. If an increase in error signal is detected, ANC is deactivated and the primary path is re-estimated, followed by the secondary path being remodeled using the control signal. However, this procedure complicates the system and makes its implementation more challenging.  

In this paper, we propose a mode-switching strategy to enhance the modified FXLMS algorithm with online modeling of secondary paths. The proposed system alternates between adaptive ANC and online SPM without the use of additional filters or noise. Initialization is performed as usual, then the adaptive ANC is operated using a modified FXLMS algorithm. When a secondary path change is determined to be the cause of the divergence, the system performs online SPM with the control signal while ANC is controlled by a fixed control filter. The system will revert to adaptive ANC mode as soon as the remodeled secondary path is complete.

The structure of the remaining sections of this paper is as follows: Section \ref{sec:method} details the proposed online SPM based on the method of mode switching. In Section \ref{sec:simulation}, several numerical simulations are performed to demonstrate the validity of the proposed method. The conclusion is presented in Section \ref{sec:conclusion}.

\section{METHODOLOGY}
\label{sec:method}
\noindent
This section first reviews the modified FXLMS algorithm in Section 2.1, followed by an introduction of the online secondary path modelling algorithm based on this structure in Section 2.2. Finally, the method for switching modes between adaptive ANC and online secondary path modelling is described in Section 2.3.

\subsection{Modified FXLMS Algorithm}
\label{ssec:MFXLMS}
\begin{figure}[h!]
\begin{center}
  \includegraphics[width=12cm]{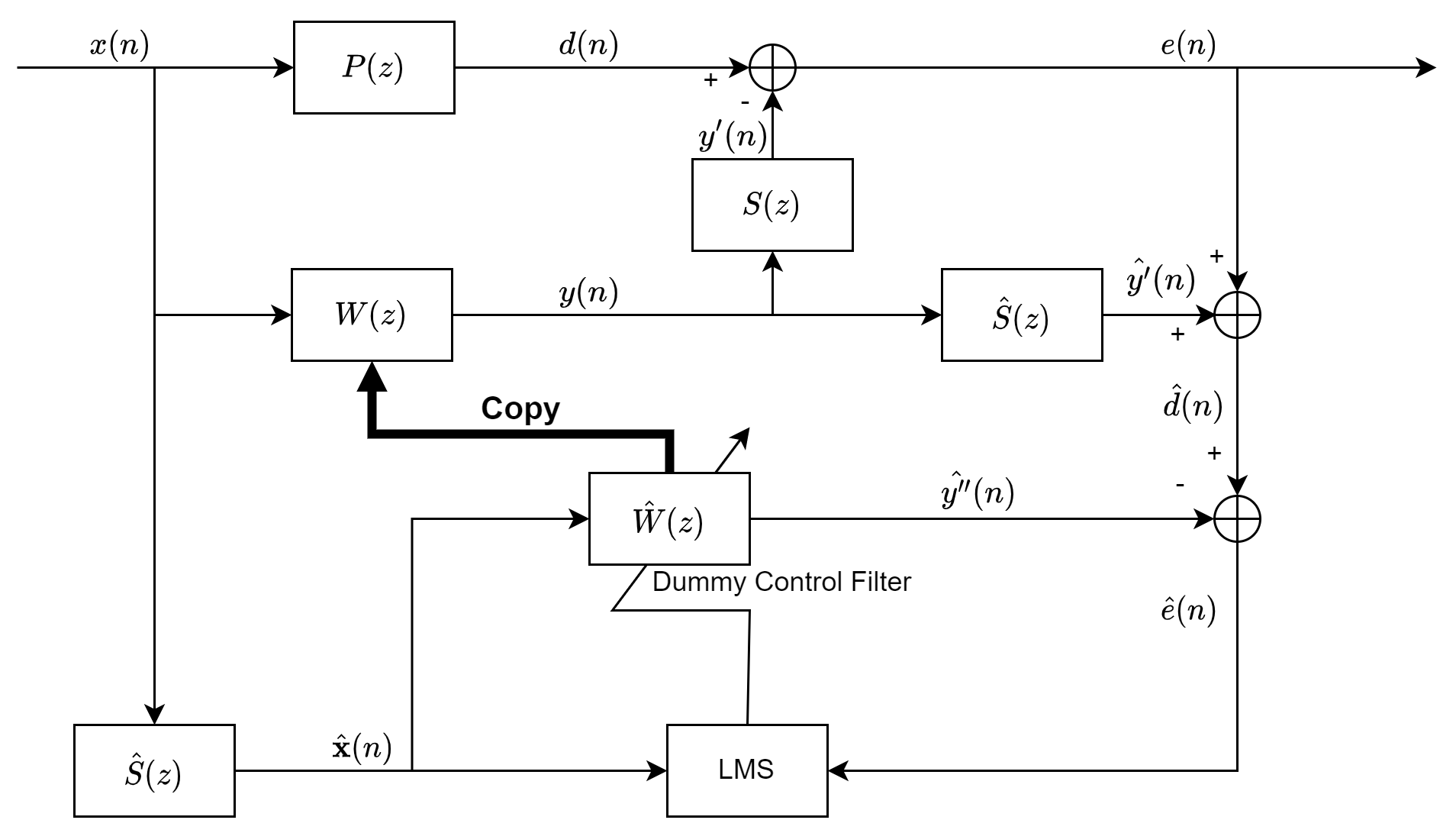}
  \end{center}
  \caption{The block diagram of modified FXLMS algorithm}
  \label{fig:1 MFXLMS}
\end{figure}

\noindent
The modified filtered X least mean square (FXLMS) algorithm is widely used due to its rapid adaption of the controller \cite{BjarnasonMFXLMS}. The block diagram is shown in Figure \ref{fig:1 MFXLMS}, in which a dummy adaptive control filter, $\hat{\mathbf{w}}(n)$, is used to suppress the estimate of disturbance signal, $\hat{d}(n)$. Then the coefficients of the dummy adaptive filter are copied to the controller, $\mathbf{w}(n)$, to generate the control signal $y(n)$. Thus, the residual error $e(n)$ can be obtained by measuring the difference between disturbance signal $d(n)$ and the control signal received at the error sensor that is expressed as:

\begin{equation}
    e(n) = d(n) - y(n)*s(n),
    \label{eq1}
\end{equation}
\noindent
where $x(n)$ is the reference signal and $s(n)$ is the actual secondary path. '$*$' denotes convolution operation. Suppose that the secondary path can be modelled as $\hat{\mathbf{s}}(n)$, the estimated disturbance signal is given by:
\begin{equation}
    \hat{d}(n) = e(n) + y(n)*\hat{s}(n).
    \label{eq2}
\end{equation}
\noindent
Then the estimated residual error signal $\hat{e}(n)$ can be defined as:
\begin{equation}
    \hat{e}(n) = \hat{d}(n) - \hat{x}(n)*\hat{w}(n),
    \label{eq3}
\end{equation}
\noindent
where $\hat{x}(n)$ is the filtered reference signal which is calculated by passing the reference signal $x(n)$ through the estimated secondary path $\hat{s}(n)$. The dummy control filter can now be updated using LMS algorithm:
\begin{equation}
    \hat{\mathbf{w}}(n+1) = \hat{\mathbf{w}}(n) + \mu_w\hat{\mathbf{x}}(n)\hat{e}(n),
    \label{eq4}
\end{equation}
where $\hat{\mathbf{x}}(n)=[\hat{x}(n),\hat{x}(n-1),...,\hat{x}(n-L_w+1)]$ and $L_w$ is the length of control filter. $\mu_w$ represents the step size. Assuming that the control filter adjusts slowly, indicating that the time variation in the control filter can be ignored, the dummy control filter will be the same as the actual one, i.e. $\hat{w}(n) = w(n)$. Therefore, Equation \ref{eq3} becomes:
\begin{equation}
    \hat{e}(n) = \hat{d}(n) - {x}(n)*{w}(n)*\hat{s}(n).
    \label{eq5}
\end{equation}
\noindent
Replacing $\hat{d}(n)$ with Equation \ref{eq2}, the estimated residual error signal is equal to the actual one, which is described in Equation \ref{eq1}. As a result, the performance is exactly the same as the normal filtered X LMS (FXLMS) algorithm.

\subsection{Online Secondary Path Modelling}
\begin{figure}[h!]
\begin{center}
  \includegraphics[width=12cm]{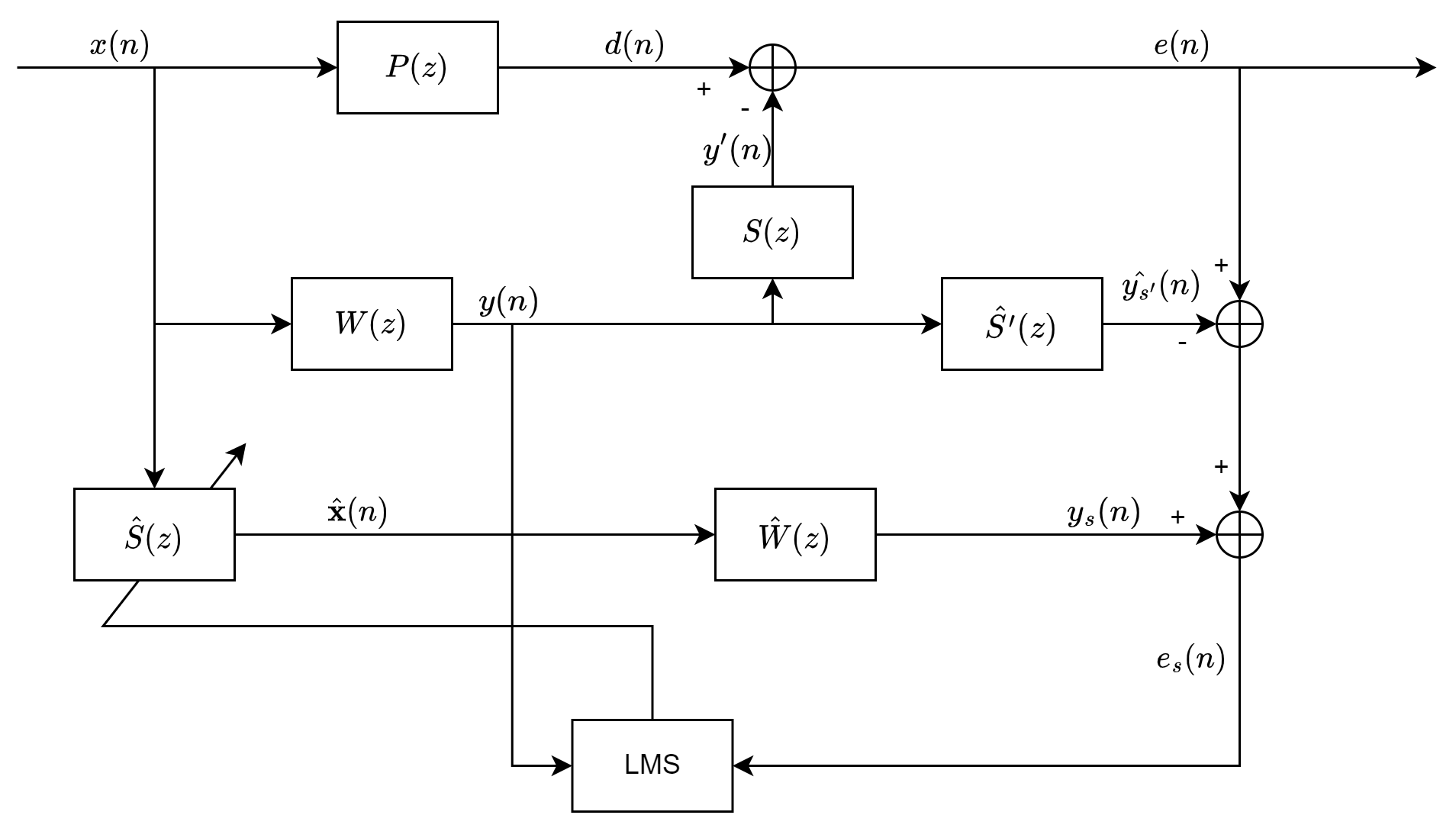}
  \end{center}
  \caption{The block diagram of online secondary path modelling}
  \label{fig:2 SPM}
\end{figure}

\noindent
Since the modified FXLMS algorithm is introduced to remove the delay effect from the secondary path, the convergence speed can thus be increased by using a large step size \cite{HansenACNV}. According to Equation \ref{eq2} that demonstrates the removal of the secondary path effect, the prerequisite is that the secondary path is well-modelled \cite{BjarAnalyfxlms, LopesBehMFXLMS}. However, the secondary path is time-variant which introduces instability into the system and further degrades the performance of the modified FXLMS algorithm which is sensitive to the modelling error of the secondary path. For the purpose of enhancing the robustness of the modified FXLMS algorithm, we propose an online secondary path modelling (SPM) method whose diagram is shown in Figure \ref{fig:2 SPM}.

The structure is slightly changed based on the modified FXLMS that is depicted in Figure \ref{fig:1 MFXLMS}, resulting in no additional filters or auxiliary noise introduced into the system. In the procedure, the inner error signal for modelling the secondary path should be:
\begin{equation}
    e_s(n) = d(n)-y(n)*s(n)-y(n)*\hat{s}'(n)+x(n)*\hat{s}(n)*\hat{w}(n),
    \label{eq6}
\end{equation}
\noindent
where $\hat{s}'(n)$ is the previous estimated secondary path and $\hat{w}(n)$ is exactly equal to the control filter $w(n)$. The Equation \ref{eq6} can be further derived to:
\begin{equation}
    e_s(n) = d(n)-y(n)*s(n)-y(n)*\hat{s}'(n)+y(n)*\hat{s}(n).
    \label{eq7}
\end{equation}
\noindent
Thus, the instantaneous cost function is:
\begin{equation}
    J(n) = (e_s(n))^2,
    \label{eq8}
\end{equation}
\noindent
In order to minimize Equation \ref{eq8}, its negative gradient in terms of $\hat{\mathbf{s}}(n)$ is used to update the estimated secondary path vector as:
\begin{equation}
    \hat{\mathbf{s}}(n+1) = \hat{\mathbf{s}}(n) - \mu_s\mathbf{y}(n)e_s(n),
    \label{eq9}
\end{equation}
\noindent
where $\mu_s$ is the step size and $\mathbf{y}(n) = [y(n), y(n-1),...,y(n-L_s+1)]$ is the control signal vector with the length of the estimated secondary path, $L_s$. Noting that if the primary path is stable and the ANC is converged, the signal $\hat{y_s}'(n)$ that is obtained by $y(n)*\hat{s}'(n)$ can be considered to have similar value as disturbance signal $d(n)$. Consequently, Equation \ref{eq7} reflects the difference between the actual and estimated secondary path.

The proposed online SPM approach enables the system to separate the functions of secondary path modelling and noise reduction. More specifically, during modelling the secondary path, the control signal $y(n)$ is generated by a fixed control filter, contributing to maintaining certain noise reduction performance. 

\subsection{Mode Switching Method}
\begin{figure}[h!]
\begin{center}
  \includegraphics[width=8cm]{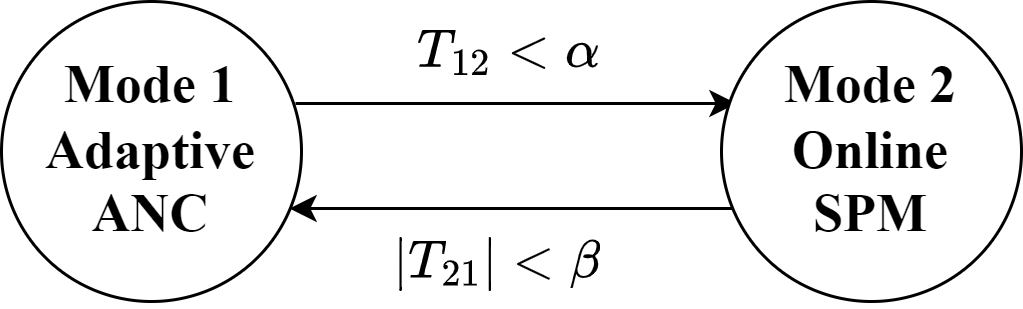}
  \end{center}
  \caption{The transfer diagram of mode switching between adaptive ANC and online SPM}
  \label{fig:3 modeswitch}
\end{figure}

\noindent
In order to improve the modified FXLMS algorithm with online secondary path modelling while less computational cost is increased, we propose a mode switching method, where the system switches the operating mode between adaptive ANC and online SPM as depicted in Figure \ref{fig:3 modeswitch}.

In mode 1, the system does adaptive ANC using a modified FXLMS algorithm. When the secondary path change is detected, manifesting the phenomenon of divergence, the system switches to mode 2 to remodel the secondary path. To monitor the divergence, the judgement criterion $T_{12}$ in Figure \ref{fig:3 modeswitch} is defined as:
\begin{equation}
    T_{12}=10log_{10}(\frac{P_x(n)}{P_{e}(n)}),
    \label{eq10}
\end{equation}
\noindent
where $P_x(n)$ and $P_e(n)$ are the instantaneous power of reference signal $x(n)$ and residual error $e(n)$, respectively. They can be calculated as the following recursive equations \cite{Pradhan5stage}:
\begin{equation}
    P_{x}(n)=\lambda P_{x}(n-1)+(1-\lambda)x^2(n),
    \label{eq11}
\end{equation}
\noindent
\begin{equation}
    P_{e}(n)=\lambda P_{e}(n-1)+(1-\lambda)e^2(n),
    \label{eq12}
\end{equation}
\noindent
where $\lambda$ is the forgetting factor, ranging from 0.9 to 1. If the secondary path changes, the residual error signal becomes divergent, leading to a high power of the residual error that will decrease the value of $T_{12}$. The ratio of the power of the reference signal to the power of the residual error signal can thus well reflect the noise reduction performance of the system without considering the variation of the reference signal due to that $x(n)$ and $e(n)$ are linearly related.

At the point when $T_{12}$ is smaller than the threshold value $\alpha$, the system will operate in mode 2 to model the changed secondary path. To detect the moment that the changed secondary path is remodelled, the judgement criterion  $T_{21}$ depending on the slope of modelling error signal $e_s(n)$ can be described as:
\begin{equation}
    T_{21}=\frac{\partial [E(e_s^2(n))]}{\partial n}=E[2e_s(n)\frac{\partial e_s(n)}{\partial n}].
    \label{eq13}
\end{equation}
\noindent
Note that Equation \ref{eq13} can be further derived in a more practical expression by substituting the expectation and derivation with the time average and first difference \cite{ShenAltern}, which is:
\begin{equation}
    T_{21}=\frac{1}{N} \sum_{i=0}^{N-1} e_s(n-i)[e_s(n-i)-e_s(n-i-1)],
    \label{eq14}
\end{equation}
\noindent
where $N$ is the number of average samples, being used to smooth the slope. It is obvious that if the slope of the learning curve approaches zero indicating that the secondary path modelling error signal $e_s(n)$ become small, the changed secondary path is well modelled. In a real situation, the slope is hard to get zero value. We set a threshold $\beta$ to determine the time to switch back to adaptive ANC mode, i.e. $|T_{21}|$ is smaller than $\beta$.

Overall, the system starts from mode 1 and adaptively eliminates unwanted noise until the secondary path changes and is detected. Then, it switches to mode 2 to estimate the changed secondary path while the control signal for suppressing the noise is generated by the fixed control filter, ensuring a certain noise reduction level. Once the secondary path is remodelled, the system returns to adaptive ANC mode. Therefore, the proposed method degrades the modified FXLMS algorithm's sensitivity to the secondary path by online SPM, and the mode switching method not only does not increase the computational complexity but also maintains certain ANC's performance.

\section{SIMULATION}
\label{sec:simulation}
\begin{figure}[h!]
\begin{center}
  \includegraphics[width=12cm]{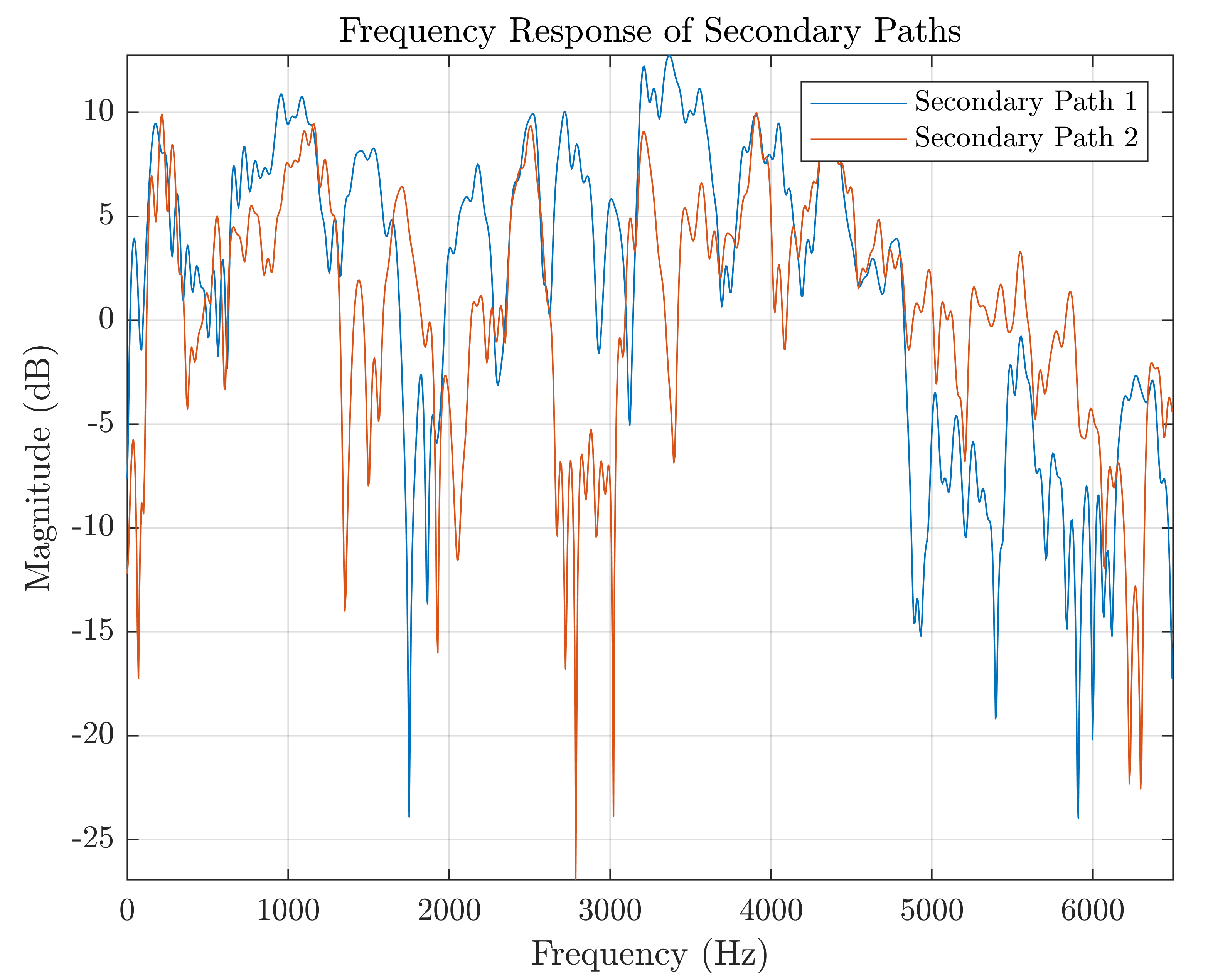}
  \end{center}
  \caption{Frequency response of two secondary paths}
  \label{fig:4 secondarypath}
\end{figure}

\noindent
In this section, we validated the proposed online SPM with mode switching ANC system through numerical simulation. The primary noise is a broadband signal with a frequency range of $100$ to $1000$ Hz and the sampling frequency is 13000Hz. The primary path is band-pass filters with a frequency band between $80$ and $5000$ Hz and the control filter consists of $512$ taps. The two secondary paths are measured from a real environment with $256$ FIR filter tap length that is shown in Figure \ref{fig:4 secondarypath}. The threshold $\alpha$ and $\beta$ are selected as $10$ and $0.0000001$ respectively and the forgetting factor $\lambda$ is equal to $0.999$. The step size $\mu_w$ and $\mu_s$ are chosen as $0.0001$ and $0.001$ while $N$ is $64$ for smoothing the slope. To quantify performance, the mean square error (MSE) is defined as:
\begin{equation}
    \mathrm{MSE} = 10log_{10}(E(e^2(n))),
    \label{eq15}
\end{equation}
\noindent

 Simulations have also been carried out to compare the performance with 5 stage method \cite{Pradhan5stage}, where the parameters are set as $T_{r0} = 10$, $T_{p0} = 20$, $T_{s0} = 20$, $\mu_p = 0.00001$, $\mu_w = 0.0001$ and $\mu_s=0.001$, and Akhtar's method \cite{AkhtarVSSSPM}, in which $\mu_w = 0.001$, $\mu_s = 0.001$ $\mu_{s_{min}} = 0.0001$, $\mu_{s_{max}} = 0.001$ and the injected noise is white Gaussian with a variance of 0.001.

\subsection{Case 1}
\begin{figure}[h!]
\centering
    \begin{subfigure}{8cm}
        \includegraphics[width=8cm]{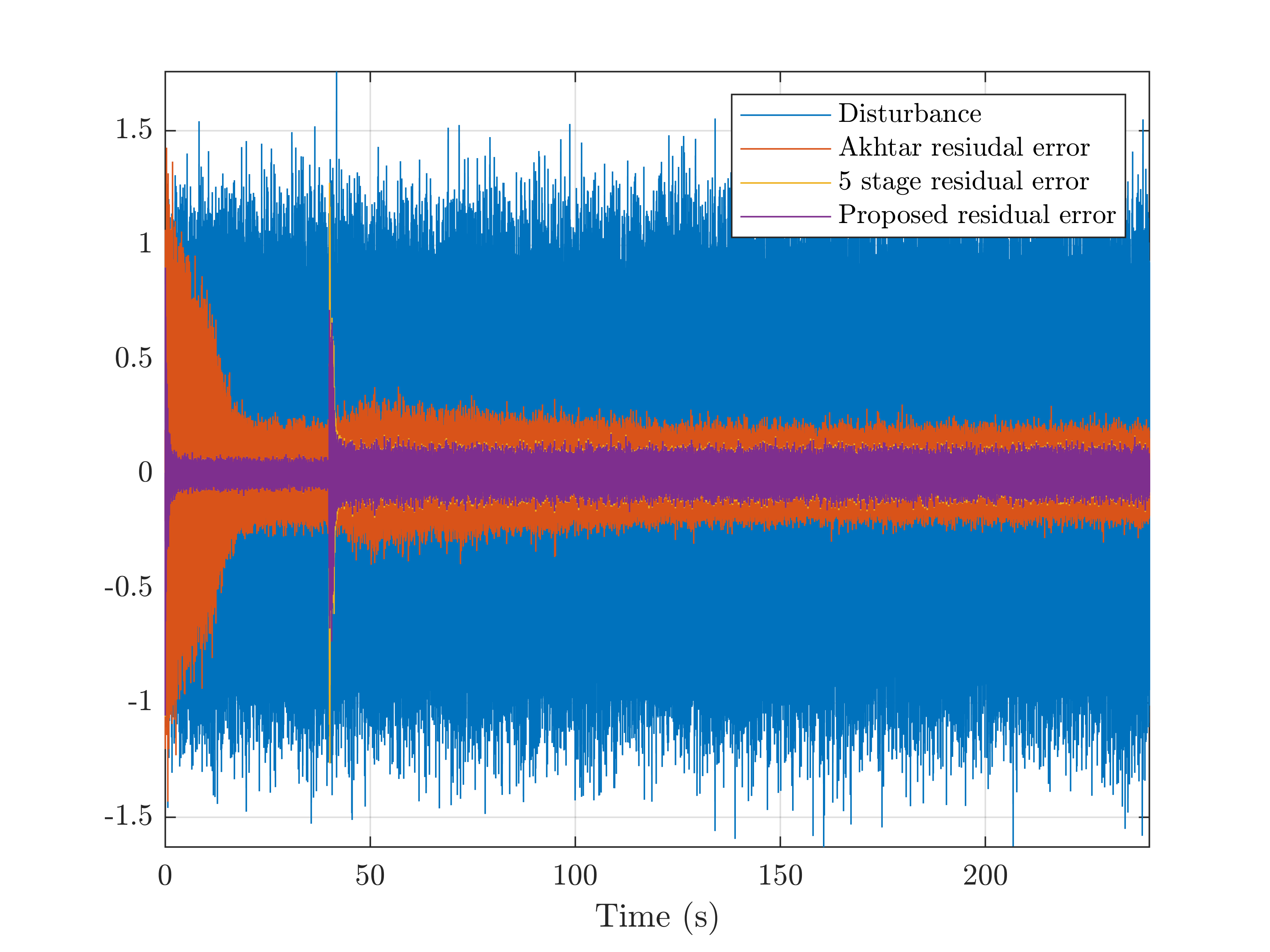}
        \label{fig:5(a)}
    \end{subfigure}
    \begin{subfigure}{8cm}
        \includegraphics[width=8cm]{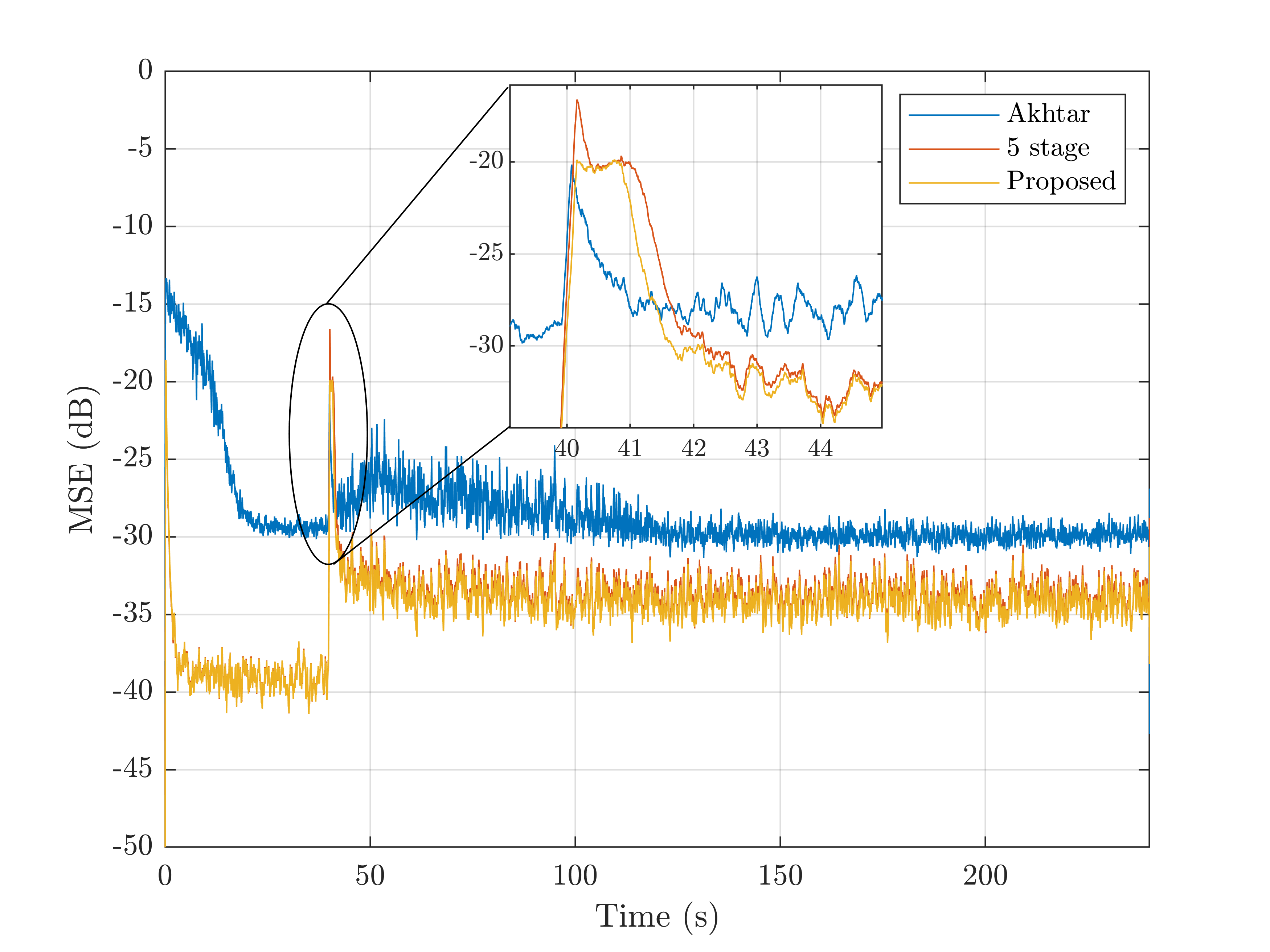}
        \label{fig:5(b)mse}
    \end{subfigure}
    \caption{(a) Case 1 residual error signal (b) Case 1 learning curve}
    \label{fig:5 simulation result}
\end{figure}

\noindent
In this case, we validate the performance of the proposed method where the secondary path is changed from 'Secondary Path 1' to ' Secondary Path 2' at the $40$th second. As shown in Figure \ref{fig:5 simulation result} that all the algorithms can react to changes in the secondary path and then remodel it. Both 5 stage method and the proposed method have almost identical noise reduction levels and also have better performance than Akhtar's method in the steady state. It can also be seen from Figure \ref{fig:5 simulation result}b that although Akhtar's method has a faster convergence rate after secondary path changes compared to the other two methods, it introduces additional noise making the ANC noise reduction less effective than those two methods. The relatively flat part of MSE in the sub-figure of Figure \ref{fig:5 simulation result}b is the proposed system working in mode 2, for secondary path remodelling. It is also noted that the 5 stage approach has a spike in the figure, this is because it requires a process of re-estimating the primary path before remodelling the secondary path, where the ANC is not working. The method proposed in this paper instead ensures that the ANC works as an adaptive or fixed filter all the time, keeping the noise reduction performance at a certain value, i.e. the threshold $\alpha$, resulting in a better noise reduction performance.

\begin{figure}[h!]
\begin{center}
  \includegraphics[width=7.5cm]{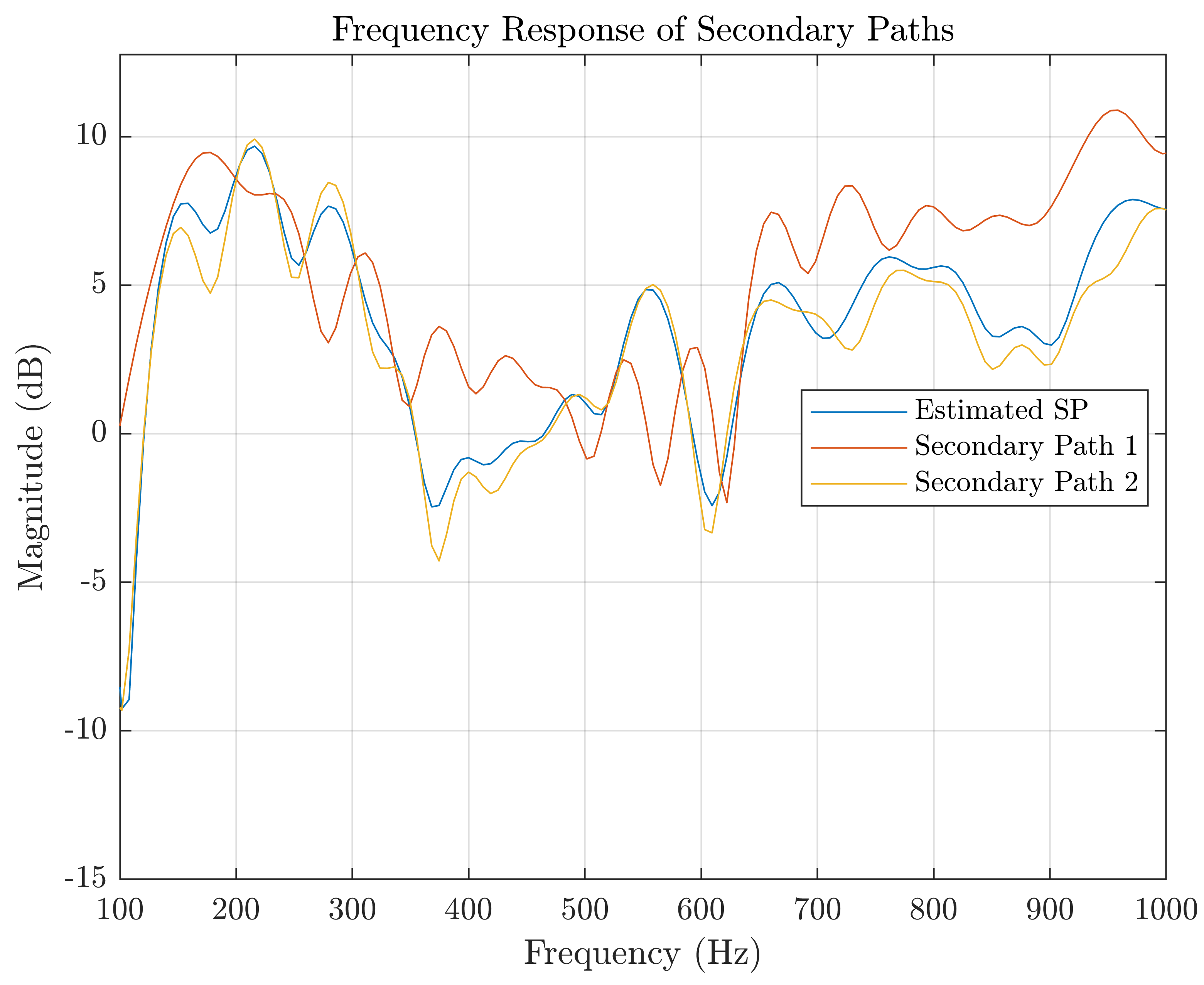}
  \end{center}
  \caption{Frequency response of estimated secondary path after being remodelled comparing with two secondary paths}
  \label{fig:6 estimatedsecondarypath}
\end{figure}

The frequency response shown in Figure \ref{fig:6 estimatedsecondarypath} also illustrates that the proposed method can effectively remodel the changed secondary path in the frequency band between 100Hz and 1000Hz due to the control signal being a broadband signal within that band. The frequency band in which the secondary path can be modeled depends on the frequency range of the control signal.

\subsection{Case 2}
\begin{figure}[h!]
\centering
    \begin{subfigure}{8cm}
        \includegraphics[width=8cm]{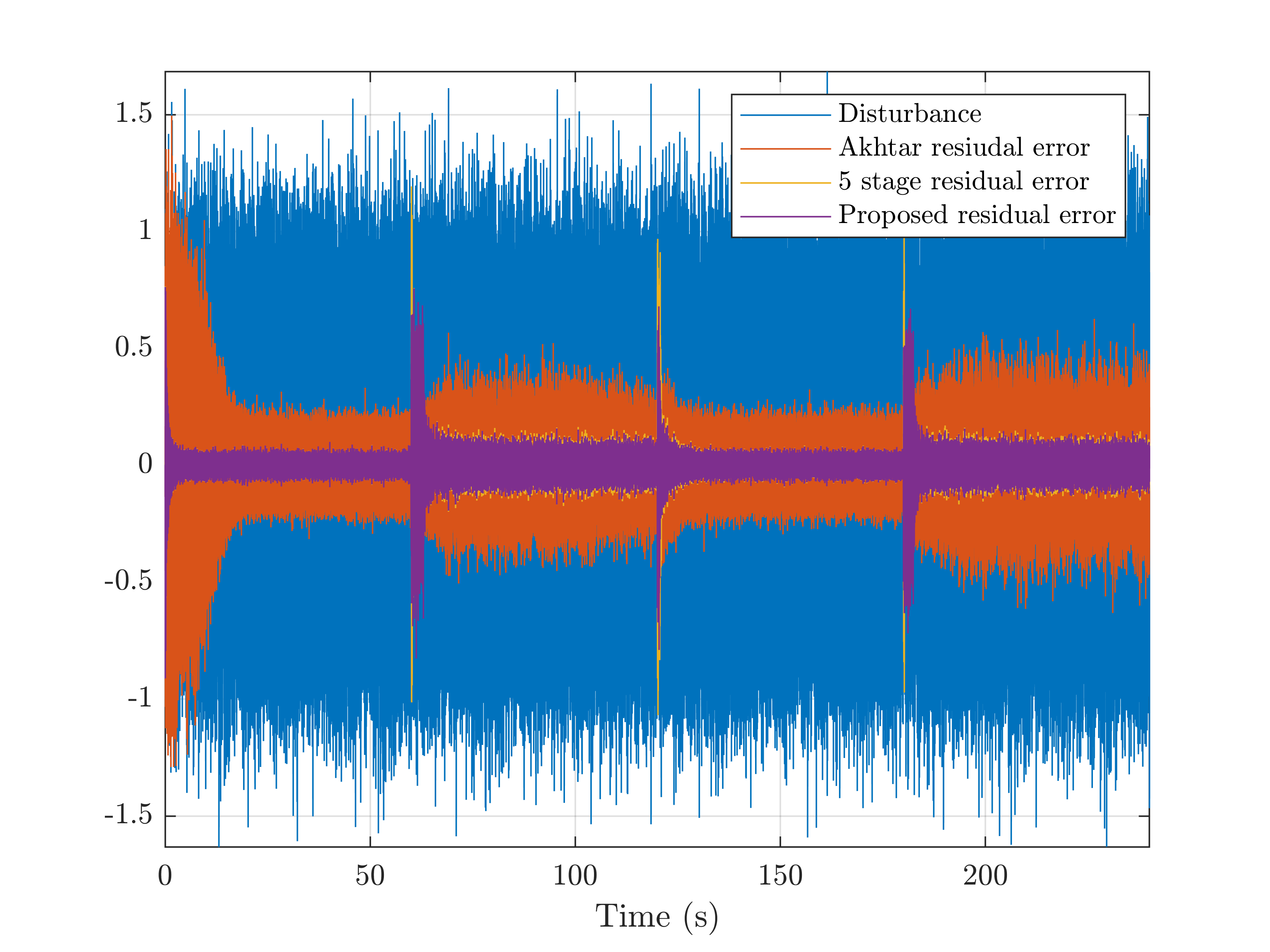}
        \label{fig:7(a)}
    \end{subfigure}
    \begin{subfigure}{8cm}
        \includegraphics[width=8cm]{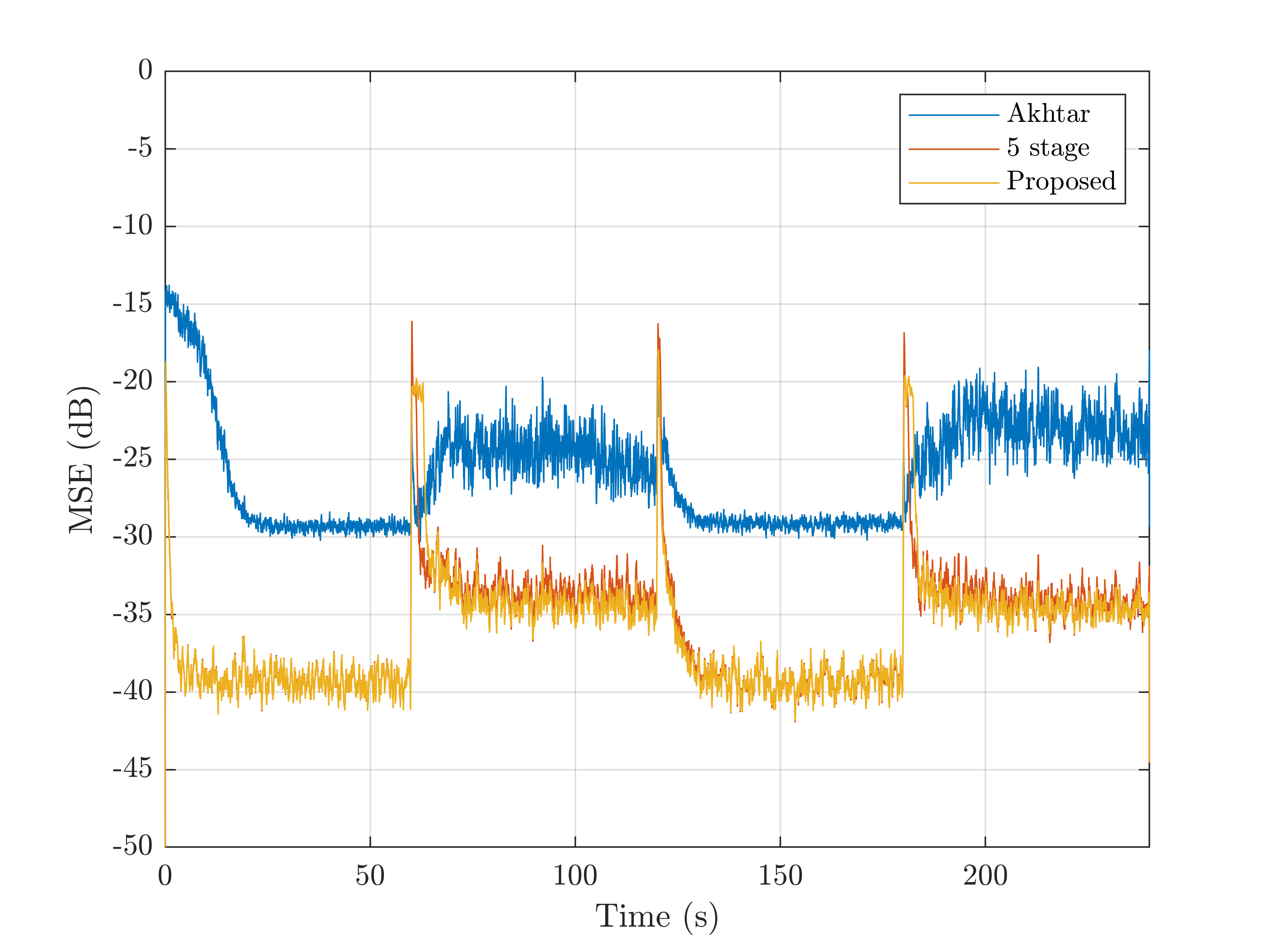}
        \label{fig:7(b)}
    \end{subfigure}
    \caption{(a) Case 2 residual error signal (b) Case 2 learning curve}
    \label{fig:7 simulation result}
\end{figure}

\noindent
In this case, we investigate that the secondary path is changed multiple times. It changes the actual secondary path every minute, switching back and forth between "Secondary Path 1" and "Secondary Path 2" to simulate frequent path changes. It can be seen in Figure \ref{fig:7 simulation result} that 5 stage and the proposed method can both cope with frequent changes in the secondary path and end up with nearly identical noise reduction, which are all better than Akhtar's method. However, as mentioned in the previous section, the 5-stage method stops the ANC when the primary path is re-estimated each time when path changes are detected, which is not as effective as the proposed system where the ANC is always working. 

\section{CONCLUSIONS}
\label{sec:conclusion}
\noindent
This paper presents an online secondary path modelling approach to improve the modified FXLMS algorithm's sensitivity to the secondary path. By means of a mode switching approach, the system switches between adaptive ANC and online SPM. The proposed system can quickly respond to the divergence caused by changes in the secondary path and remodel it. In contrast to existing methods, the proposed method, in which the ANC operates as an adaptive or fixed control filter during any mode, ensures that the system has a certain level of noise reduction at all times.

%
\section*{Acknowledgements}
\noindent
This research is supported by the Singapore Ministry of Education, Academic Research Fund Tier 2, under research grant MOE-T2EP50122-0018


\bibliographystyle{unsrt}
\bibliography{sample} 

\end{document}